\documentclass[aps,pra,twocolumn,floatfix,superscriptaddress,times]{revtex4}
\usepackage{amsmath,amssymb,amsfonts,bbm,graphicx,color,mathptmx}
\usepackage{hyperref}
\usepackage[utf8]{inputenc}
\usepackage[T1]{fontenc}

\newcommand{\ket}[1]{\vert #1 \rangle}
\newcommand{\bra}[1]{\langle #1 \vert}

\def\i{{rm i}}

\def\p{{\cal P}}
\begin{document}
\title{Hybrid quantum key distribution using coherent states 
and photon-number-resolving detectors}
\author{Marco Cattaneo}
\affiliation{Quantum Technology Lab, 
Dipartimento di Fisica ``Aldo Pontremoli'', Universit\`a degli Studi di
Milano, I-20133 Milano, Italy}
\author{Matteo G. A. Paris}
\affiliation{Quantum Technology Lab, 
Dipartimento di Fisica ``Aldo Pontremoli'', Universit\`a degli Studi di
Milano, I-20133 Milano, Italy}
\affiliation{INFN, Sezione di Milano, I-20133 Milano, Italy}
\author{Stefano Olivares}
\email{stefano.olivares@fisica.unimi.it}
\affiliation{Quantum Technology Lab, 
Dipartimento di Fisica ``Aldo Pontremoli'', Universit\`a degli Studi di
Milano, I-20133 Milano, Italy}
\affiliation{INFN, Sezione di Milano, I-20133 Milano, Italy}
\date{\today}
\begin{abstract}
We put forward a hybrid quantum key distribution protocol based on 
coherent states, Gaussian modulation, and photon-number-resolving (PNR) 
detectors, and show that it may enhance the secret key generation rate 
(KGR) compared to homodyne-based schemes. Improvement in the 
KGR may be traced back to the dependence of the 
two-dimensional discrete output variable on {\em both} the input quadratures, thus overcoming 
the limitations of the original protocol.
When reverse reconciliation is considered, the scheme based
on PNR detectors outperforms the homodyne one both for individual and collective
attacks. In the presence of direct reconciliation, the PNR strategy is still
the best one against individual attacks, but for the collective ones the homodyne-based
scheme is still to be preferred as the channel transmissivity decreases.
\end{abstract}
\maketitle
\section{Introduction}
In the last decade, continuous-variable quantum key distribution (QKD) 
based on coherent states and homodyne detection (HD-QKD)  
gained much attention in the cryptographic community \cite{diamanti:npj:16}.
In particular, the compatibility with telecom techniques and the high-detection
efficiency makes HD-QKD of interest for practical implementations \cite{tob15}.
{Moreover, the recent advances in establishing continuous-variable
ground-satellite quantum channels \cite{leuchs:17,vcu17} exploiting coherent states 
and homodyne detection has opened the way to the possibility of a global 
QKD network.} Heterodyne based protocols has been also suggested \cite{het1} 
and the secret key rate valid against individual attacks has been analyzed \cite{het2,het3}.
\par
Usually, in HD-QKD an observable with a continuous spectrum is used to
encode the information which will be used to extract the secret key. For example,
in the original continuous-variable (CV) protocol based on coherent states 
\cite{Grosshans2001,gr03}, the information is encoded by Gaussian modulation of
phase and amplitude of an input coherent state, whereas the secret key is
retrieved by a slicing protocol processing the data from a homodyne detector.
\par
Here we consider a CV-QKD protocol based on coherent states, but we substitute 
the homodyne detector with a scheme based on photon-number-resolving (PNR) 
detectors. While in a typical homodyne detection one measures the difference
photocurrent from a couple of pin photodiodes \cite{oli:PRA:2013} able to 
detect a macroscopic photocurrent proportional to the number of photons, 
here we consider a scenario in which the {\it number} of photons is measured 
at the two detectors. Recent theoretical
\cite{vogel:PRA:93,sperling:EPL:2015} and experimental results \cite{bina:OE:2017,BinaOlivares}
have shown that detection schemes aimed to measure the photon number statistics can be
exploited in order to obtain some useful information about the field quadratures. Motivated
by these results, we investigate whether, and to which extent, these 
PNR-detection schemes can be employed in QKD protocols. Throughout the paper 
we will refer to this kind of protocol as
PNR-QKD. More in details, we will address the mutual information 
between sender, receiver and eavesdropper, as a figure of merit 
to assess the performance of the PNR-based protocols \cite{maurer1993secret}.
\par
The paper is structured as follows. In section~\ref{s:HD:QKD} we review the
principles of the HD-QKD based on coherent states and evaluate the mutual
information between sender and receiver and between sender and eavesdropper, 
in order to assess the maximum key generation rate (KGR) \cite{maurer1993secret}. Then, section~\ref{s:PNR:QKD} illustrates our novel PNR-QKD: we show that the 
presence of two PNR detectors allows to extract more information
about the detected signals.
In our analysis we consider the couple of numbers corresponding
to the detected photons as a two-dimensional statistical variable.
We find that there exists a threshold value on the LO energy above which
PNR-QKD outperforms HD-QKD. In this regime,  we investigate the performance of
the PNR-QKD with respect to HD-QKD in the presence of individual and
collective attacks and for direct and reverse reconciliation.
Section~\ref{s:concl} closes the paper with some concluding remarks.
\section{HD-QKD with coherent states}\label{s:HD:QKD}
\begin{figure}[tb]
\centering
\includegraphics[width=0.9\columnwidth]{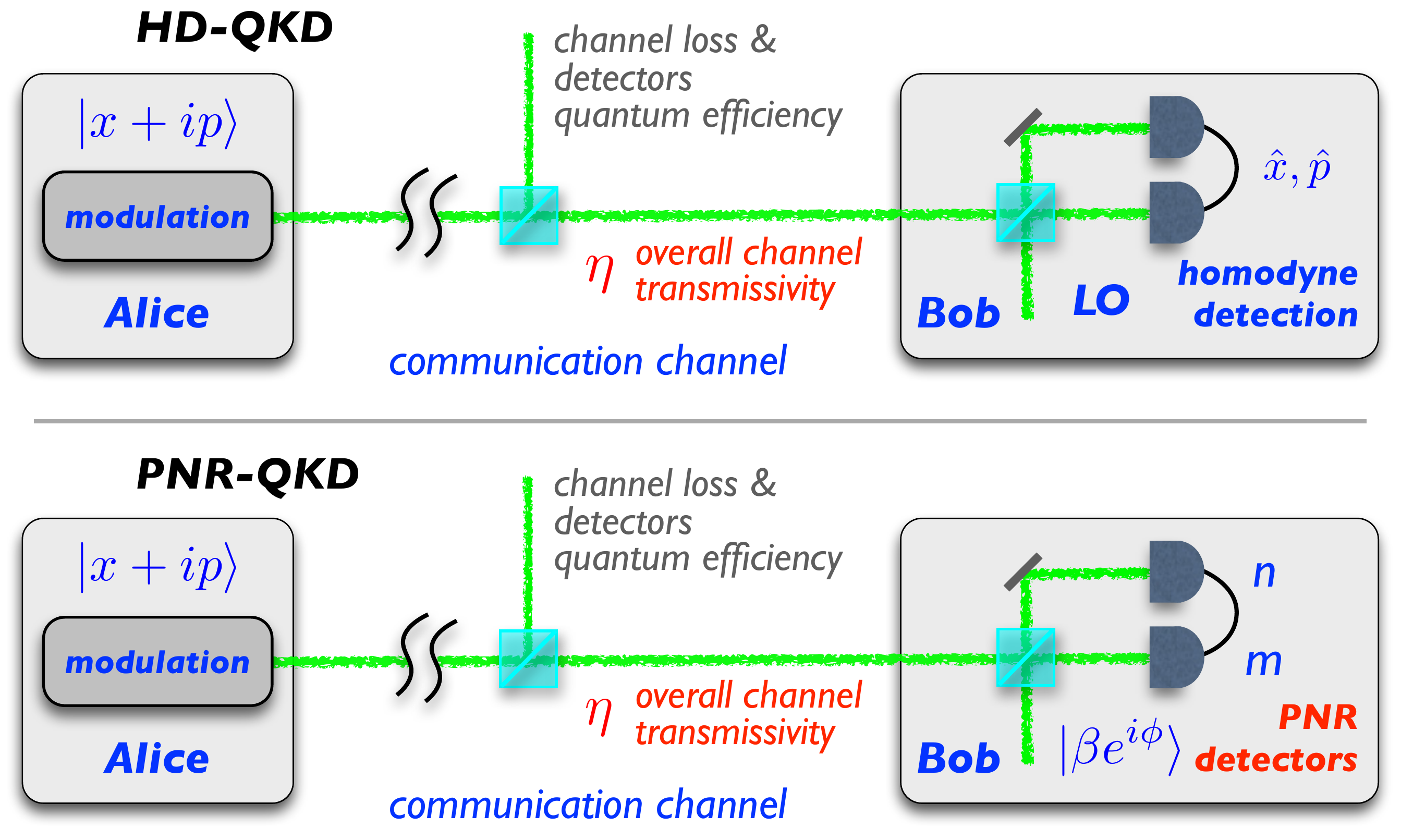} 
\vspace{-0.3cm}
\caption{(Top) Scheme of homodyne detection-based QKD (HD-QKD).
(Bottom) Scheme of PNR-based QKD, in which the two PNR detectors are
used together with a low-intensity LO. The parameter $\eta$ refers to the
overall channel transmissivity.} See the text for details.
\label{f:scheme}
\end{figure}

In HD-QKD (top panel of Fig.~\ref{f:scheme}), Alice, the sender,
draws two random real numbers, $x$ and $y$, from a normal 
distribution ${\cal N}_{\mu,\sigma^2}(z)$, with mean value
$\mu = 0$ and variance $\sigma^2=\Sigma^2$.
Then, Alice prepares the coherent state $\ket{\alpha} =\ket{x+iy}$,
which is sent to Bob through a quantum channel. Then, the receiver,
Bob performs homodyne detection by mixing the
signal at a balanced beam splitter (BS) with a local oscillator (LO),
i.e. a highly excited coherent state. Upon setting the LO phase at
either $\phi=0$ or $\phi=\pi/2$, Bob can detect the quadratures 
$\hat{x}$ or $\hat{y}$, respectively. In order to distribute a secret 
key, Bob chooses randomly to measure either one quadrature or 
the other on the signal received from Alice. After repeating this 
procedure several times, the partners share a string of (real) 
random variables, whose correlations are quantified by the mutual 
information $I(A;B)$, where $A$ and $B$ are the random variables 
of Alice and Bob, respectively, with joint distribution 
$P_{AB}(a,b)$ and marginal distributions $P_{A}(a)$ and $P_{B}(b)$.
\par
In the ideal case of lossless channel, we have \cite{Grosshans2001}:
\begin{align}
I(A;B) &= H(A;B) - H(A|B) - H(B|A), \\
&=\frac{1}{2}\log_2(1+4\Sigma^2), \label{eqn:iabid}
\end{align}
where $H(A;B)$ is the joint (Shannon) entropy, $H(A|B)$ and $H(B|A)$ 
are the conditional entropies of $A$ and $B$.
\par
Losses, including the quantum efficiency of the detectors,
can be described by an \emph{overall channel  transmissivity} $\eta$, $0\leq\eta\leq 1$
(see Fig.~\ref{f:scheme}).
In this case, the mutual information between the parties is unavoidably reduced 
and becomes:
\begin{equation}
\label{eqn:iabconeve}
I(A;B)=\frac{1}{2}\log_2(1+4\eta\Sigma^2).
\end{equation}
Starting from their shared correlated variables, 
Alice and Bob may distill a secret shared  key using \textit{reconciliation}
\cite{slicedrec} and  \textit{privacy amplification} \cite{Bennett1995Gen},
both making  use of a classical channel. 
The presence of losses allows an eavesdropper, Eve,
to obtain some information  about Alice's random variable without being detected,
upon hiding herself within the channel loss.
In this case, the amount of information is limited by the no-cloning theorem
\cite{cerf2000optimal,grosshans2001nocloning}.
More in details, the reconciliation stage requires a
flow of information from Alice to Bob (direct
reconciliation, DR) or viceversa (reverse reconciliation, RR)
in order to correct the transmission errors and agree on a common
bit string, which is thus partially known by the eavesdropper.
In the first case (DR), Alice's data form the secret key, therefore it is important
to evaluate the information shared between Alice and Eve. In RR, the secret key
is instead based on Bob's data and the information shared between Bob and Eve
becomes the relevant player. The evaluation of the information shared between the
parties requires to understand also what is the kind of measurement
performed. Here we focus on individual attacks (each pulse signal is measured
individually) and collective attacks (the measurement involves all the sent pulses)
\cite{GrosshanscontvarQKD}. 
In the following we assume that also Eve uses homodyne detection.
This is a standard, feasible technique which mimics the detection system used by
Bob but with a strong LO and
pin photodiods without-photon-number resolving capabilities
(for a more general discussion on the optimality of Gaussian attacks
see Refs.~\cite{optGauss} and \cite{optGauss:2}).
\par
In the presence of individual attacks, the information shared by Eve and Alice
is quantified by the corresponding mutual information
$I(A;E)=\frac{1}{2}\log_2\left[1+4(1-\eta)\Sigma^2\right]$.
Assuming DR, the secret KGR is given by \cite{maurer1993secret}:
\begin{equation}
\label{eqn:keygenrate}
\Delta I_{\rm D}^{\rm (ind)}=I(A;B)-I(A;E)\,,
\end{equation}
that is valid for \emph{any}
QKD scheme where classical communication is permitted.
In particular, upon running the homodyne-based protocol proposed 
in Ref.~\cite{Grosshans2001}, we have:
\begin{equation}
\label{eqn:rateomom}
\Delta I_{\rm D}^{\rm (ind)}=\frac{1}{2}\log_2\left[\frac{1+4\eta\Sigma^2}
{1+4(1-\eta)\Sigma^2}\right]
\end{equation}
which is positive if $\eta>0.5$, that is, the overall losses should 
be less than 3~dB.
This limit can be beaten using RR, obtaining the following
KGR :
\begin{align}
\Delta I_{\rm R}^{\rm (ind)}&=
I(A;B)-I(E;B),\\
&=\frac{1}{2}\log_2\left[\frac{1+4\Sigma^2}
{1+4(1-\eta)\Sigma^2}\right].
\label{eqn:rateomom:rev}
\end{align}
\par
When Eve can perform collective measurements (but Bob doesn't),
we should substitute to $I(A;E)$ and $I(B;E)$ the Holevo information
\cite{Holevo} in the previous formulas. The analytical results are quite cumbersome,
but they can be obtained straightforwardly given the state of Eve conditioned
to Bob's measurement outcome \cite{oli:rev} as we will describe in the next section
(see also Ref.~\cite{grosshans2005collective} for further details).

\section{PNR-QKD with coherent states}\label{s:PNR:QKD}

In this section we focus on the use of detectors able to discriminate the number of photons,
in order to investigate whether the KGR can be improved \cite{diamanti:npj:16}.
As we will see in the following, this provides additional information 
at the output, which may be used to improve the secret KGR.
In our scheme, the two photodiodes usually employed to build 
homodyne detection in HD-QKD are replaced by two PNR detectors.
Furthermore, the high-intensity LO needed to implement homodyne detection
is replaced with a relatively low-intensity (up to tens of photons) one, $\ket{\beta e^{i\phi}}$ with
$\beta \in {\mathbb R}$ (see the bottom panel of Fig.~\ref{f:scheme}). 
\par
In order to evaluate the statistics at the output, we recall that the output state
of a balanced BS fed by coherent states is factorized 
$U_{\sc BS} \ket{\alpha}\ket{\beta e^{i\phi}}=
\ket{(\alpha+\beta e^{i\phi})/\sqrt{2}}\ket{(\beta e^{i\phi}-\alpha)/\sqrt{2}}$,
$\alpha = x + i y$.
The photon statistics measured by the two PNR detectors at the BS outputs are thus given by two 
Poisson distributions
\begin{equation}
\p_k(n;\mu_k)=e^{-\mu_k} \mu_k^n / n!, \quad (k=1,2),
\end{equation}
where the average numbers of photocounts $\mu_k \equiv \mu_k(x,y,\beta,\phi,\eta)$ are given by
(we take into account the presence of the losses):
\begin{subequations}
\label{eqn:meanv}
\begin{align}
\mu_1 &=\frac{\eta (x^2+ y^2)+\beta^2}{2}+\sqrt{\eta}\beta(x\cos\phi+y\sin\phi),\\ 
\mu_2 &=\frac{\eta (x^2+ y^2)+ \beta^2}{2}-\sqrt{\eta}\beta(x\cos\phi+y\sin\phi\,).
\end{align}
\end{subequations}
It is worth noting that, since Eqs.~\eqref{eqn:meanv} depend on both $x$ and $p$, 
the PNR-QKD scheme may exploit the whole Alice's input random variables 
$\mathbf{X}=(X,Y)$, while, using homodyne detection, the information about 
one of them is traced out at each run, due to Bob's measurement choice 
\cite{Grosshans2001}.
\par
To investigate the performance of PNR-QKD with respect to
the homodyne-based scheme, we consider the two-dimensional statistical
variable corresponding to the two numbers of detected photons,
namely, the two-dimensional discrete random variable ${\mathbf L}=(N,M)$, $N,M>0$.
The two variables $N$ and $M$ are distributed according to the Poisson 
distributions mentioned above. The joint distribution of the detected
photons $n$ and $m$ is thus given by
\begin{align}
P_{{\mathbf X}{\mathbf L}}( x,y;n,m)=&
{\cal N}_{0,\Sigma^2}(x){\cal N}_{0,\Sigma^2}(y)
\p_1(n,\mu_1)\, \p_2(m,\mu_2).
\label{eqn:jointpois}
\end{align}
Starting from the joint distribution, we can calculate the two marginals
and evaluate  the mutual information
\begin{equation}
\label{eqn:iabpoissontot}
I({\mathbf X};{\mathbf L})=H({\mathbf X},{\mathbf L}) - 
H({\mathbf X}|{\mathbf L}) - H({\mathbf L}|{\mathbf X})\,.
\end{equation}

\begin{figure}[tb]
\includegraphics[width=0.7\columnwidth]{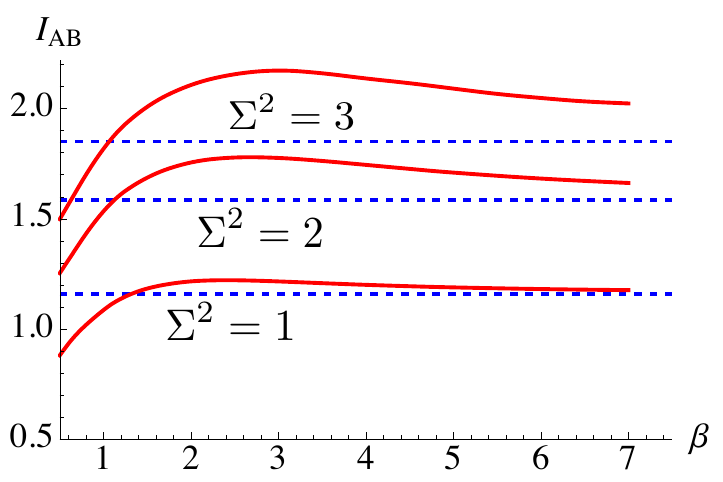} 
\vspace{-0.3cm}
\caption{The mutual information $I_{\rm AB} = I({\mathbf X};{\mathbf L})$ 
for a lossless channel as a function of $\beta$ 
and for different values of $\Sigma^2$: from bottom to top $\Sigma^2 = 1$, $2$ and $3$.
The red solid line refers to PNR-QKD and the blue
dashed one to HD-QKD (the latter is independent of $\beta$).
Notice the presence of a threshold of LO amplitude $\beta$
above which PNR-QKD outperforms HD-QKD.}
\label{fig:iabpoistot}
\end{figure}
\par
Without loss of generality we set $\phi=0$.
The mutual information for an ideal channel, with neither 
eavesdroppers nor losses, i.e. $\eta=1$, is reported in 
Fig.~\ref{fig:iabpoistot} as a function of $\beta$.
We see that there is a threshold on the value of $\beta$, 
above which the mutual information $I({\mathbf X};{\mathbf L})$
for the PNR-QKD protocol is larger than the corresponding 
quantity for the HD-QKD.
\par
\begin{figure}[tb]
\includegraphics[width=0.7\columnwidth]{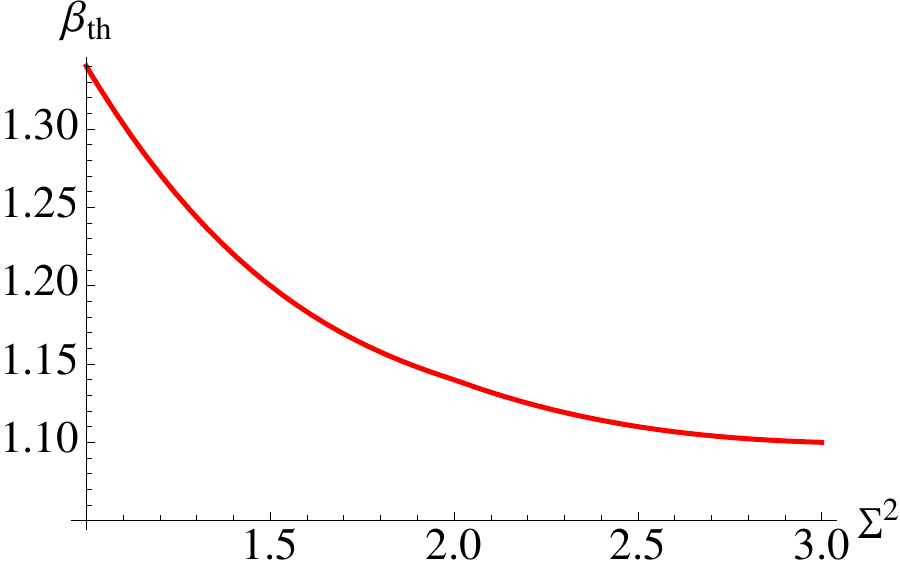} 
\vspace{-0.3cm}
\caption{Plot of the threshold value $\beta_{\rm th}$ as a function of $\Sigma^2$:
for $\beta \ge \beta_{\rm th}$ PNR-QKD outperforms HD-QKD.}
\label{fig:betath}
\end{figure}

Figure~\ref{fig:betath} shows the threshold $\beta_{\rm th}$ as a 
function of $\Sigma^2$. The monotone decreasing behaviour of $\beta_{\rm th}$
can be understand as follows.
If $\Sigma^2$ increases, we are ``feeding'' both the random variables
$X$ and $Y$ in the PNR-QKD scheme, while only one of them in HD-QKD,
i.e. we are accentuating the convenience of PNR-QKD with respect to HD-QKD.

\subsection{Individual attacks}
Let us now discuss the key generation  rate in the presence of losses and of 
an eavesdropper, Eve, in the framework of individual attacks.
In this case Eve performs a measurement in the same way on each state sent
by Alice before the reconciliation stage and the key generation
rate is given by Eq.~\eqref{eqn:keygenrate} \cite{GrosshanscontvarQKD}.
As we have seen above, the PNR-QKD 
protocol based on PNR is providing a larger information about 
the input alphabet compared to the homodyne case. Therefore,
since we have to consider the best strategy for the eavesdropper, 
we assume that Eve employs the PNR scheme, if available. 
This leads us to compare two scenarios, the one in which both Bob and Eve
employ homodyne detection and the one in which they both use PNR detectors.
The mixed case, where Bob uses homodyne detection and Eve PNR one, 
has no practical meaning, since during the reconciliation 
protocol Alice and Bob will deal only with one random variable (say $X$) and 
thus Eve's information about the other random variable $Y$ is completely 
irrelevant. This also means that Eve cannot use PNR detection in order to 
break the original homodyne protocol \cite{Grosshans2001}.
\par
\begin{figure}[tb]
\includegraphics[width=0.7\columnwidth]{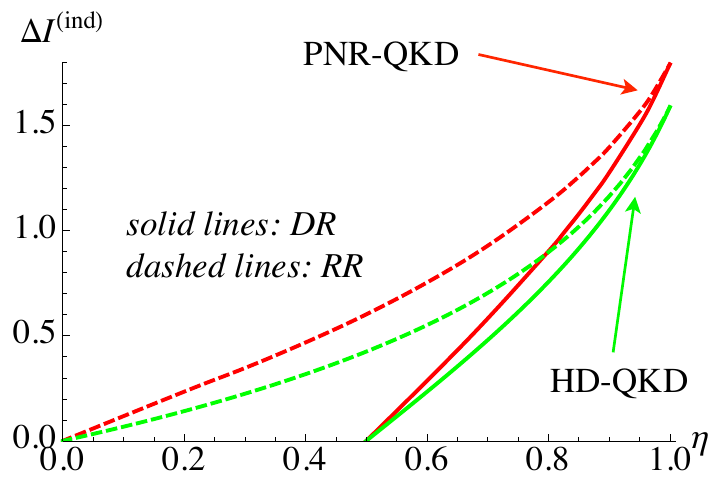}
\vspace{-0.3cm}
\caption{
The KGR $\Delta I^{\rm (ind)}$ as a function 
of the channel transmissivity $\eta$ in the presence of individual
attacks for HD-QKD (green) or PNR-QKD (red), referring to 
the scenarios in which both Bob and Eve
employ homodyne detection and the one in which 
they both use PNR detectors, respectively. We set $\beta=2$ and 
$\Sigma^2=2$. In this case PNR-QKD outperforms HD-QKD in both cases of 
DR and RR.
}
\label{fig:deltaIpois}
\end{figure}
In Fig.~\ref{fig:deltaIpois} we show the key generation 
rate $\Delta I^{\rm (ind)}$ as a function of the channel transmission $\eta$
for HD-QKD and PNR-QKD in the case of DR (solid lines) and RR (dashed lines):
PNR-QKD turns out to be the best strategy in the presence of individual attacks.
The enhancement may be traced back to the dependence of the output
variable $\mathbf{L}$ on both the full Alice's alphabet $\mathbf{X}$ and not
only on one of the components (either $x$ or $y$), as it unavoidably happens
for the original homodyne protocol. Since Bob's measurement is symmetric with
respect to Eve's one, the threshold $\eta=0.5$ does not depend on $\Sigma^2$.

\begin{figure}[tb]
\includegraphics[width=0.7\columnwidth]{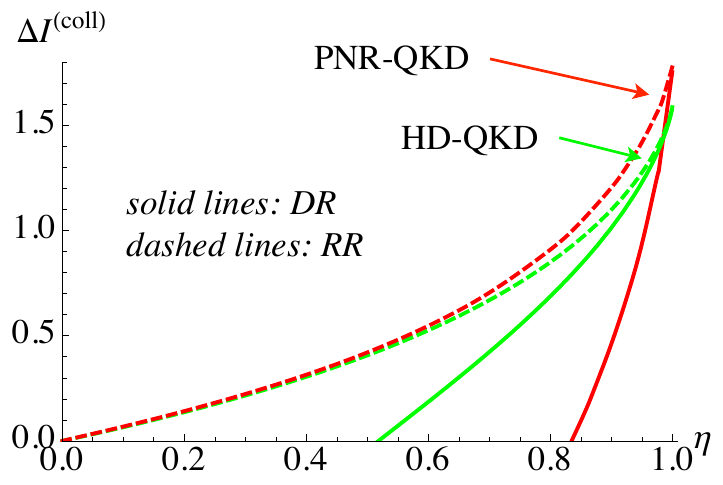}
\vspace{-0.3cm}
\caption{
The KGR $\Delta I^{\rm (coll)}$ 
as a function of the channel transmissivity $\eta$ in the presence of collective
attacks  for HD-QKD (green) or PNR-QKD (red). We set $\beta=2$ and $\Sigma^2=2$.
Now PNR-QKD beats HD-QKD for any value of $\eta$ only in the case of RR,
whereas, if we use DR, PNR-QKD turns out to be the best strategy only for high
values of the channel transmissivity.}
\label{fig:deltaI:coll}
\end{figure}
\subsection{Collective attacks}
To perform collective attacks Eve should store each quantum state
sent by Alice and measure it only after the classical key distillation procedure.
Now Eve can implement an optimal measurement, therefore the new figure of
merit for the KGR and DR is given by
(as mentioned above, for a fair comparison with the PNR-QKD we assume that
Bob does not perform a collective measurement) \cite{GrosshanscontvarQKD}:
\begin{equation}
\label{eqn:KGRcollective}
\Delta I_{\rm D}^{\rm (coll)}=I(\mathbf{X};\mathbf{L})-\chi(A;E),
\end{equation}
where $\chi(A;E)=S[\rho_E]-\sum_{\mathbf x} p({\mathbf x}) \, S[\rho_{E|{\mathbf x}}]$
is the Holevo information between Alice and Eve, with
$S[\rho] = -\text{Tr}[\rho \log_2 \rho]$ being the von Neumann entropy of the state $\rho$,
we are assuming that Eve receives the state $\rho_{E|{\mathbf x}}$ with
probability $p({\mathbf x})$ and, thus, $\rho_E=\sum_{\mathbf x} p({\mathbf x})\, \rho_{E|{\mathbf x}}$.
To calculate $\chi(A;E)$ we must distinguish HD-QKD from PNR-QKD, since in
the first case the useful information is contained only in one single quadrature (say $X$)
while in PNR-QKD in both $X$ and $Y$ (the difference appears essentially in the
reconciliation stage). According to the protocol, in the case of HD-QKD Eve receives
the mixed state (the information about the random variable $y$ is lost):
\begin{equation}
\rho_{E|x}=\int_\mathbb{R} {\cal N}_{0,\Sigma^2}(y)\,
\ket{\sqrt{1-\eta}(x+i y)}\bra{\sqrt{1-\eta}(x+i y)}\, dy
\end{equation}
with probability ${\cal N}_{0,\Sigma^2}(x)$ \cite{grosshans2005collective}.
By contrast, in PNR-QKD Alice is sending information stored in both $X$ and $Y$,
therefore Eve's conditional state is now the pure state
\begin{equation}
\rho_{E|(x,y)}=\ket{\sqrt{1-\eta}(x+i y)}\bra{\sqrt{1-\eta}(x+i y)}
\end{equation}
and $S[\rho_{E|(x,y)}] = 0$,
while $\varrho_E$
is the same as in HD-QKD.\\
\indent
When RR is considered, we should substitute
$\chi(E;B)=S[\rho_E]-\sum_{n,m} p(n,m) \, S\left[\rho_{E|(n,m)}\right]$ to $\chi(A;E)$
into Eq.~(\ref{eqn:KGRcollective}). Here $p(n,m)$ is the marginal
of $P_{{\mathbf X}{\mathbf L}}( x,y;n,m)$ given in Eq.~(\ref{eqn:jointpois}),
i.e. the joint distribution fo obtaining $n$ and $m$ number of photons at Bob's PNR detectors,
and $\rho_{E|(n,m)}$ is the corresponding conditional state received by Eve.
\\
\indent The KGR for collective attacks is depicted in
Fig.~\ref{fig:deltaI:coll} as a function of $\eta$ in the case of HD-QKD and PNR-QKD
for DR and RR. It is clear that, in the framework of collective attacks and DR (solid lines),
HD-QKD outperforms PNR-QKD (but for very high channel transmissivity!). Nevertheless,
in the presence of RR we still find that PNR-QKD beats the scheme based on homodyne
detection, though, as the channel transmissivity decreases, they exhibit almost the same
performance.

\section{Concluding remarks}\label{s:concl}
In conclusion, we have shown that PNR detectors may be profitably 
employed to design a hybrid quantum key distribution protocol using 
coherent states.
If we restrict to individual attacks, when we exploit the full information
at the output to implement PNR-QKD, the dependence of the two-dimensional
discrete output variable on {\em both} the input quadratures provides enhancement of 
the secret KGR compared to HD-QKD, both for DR and RR.
In the presence of collective attacks and RR the PNR-QKD still outperforms
the homodyne based strategy, though, as the transmission of the channel decreases,
the performance is almost the same as that of the HD-QKD. If we consider DR,
the strategy based on PNR turns out to be the best choice only for very high values
of the channel transmissivity. This is due to the conditional state received by
Eve: in the case of HD-QKD it is a mixed state (the information about one quadrature
is lost), whereas for PNR-QKD she receives pure states, having access to both
the orthogonal quadratures, and, thus leading to a clear increase of the overall
gained information.
\par
If we focus on the regimes where using PNR leads to greater $\Delta I$,
we can also note that PNR-QKD requires a lower 
value of $\eta$ with respect to HD-QKD in order to obtain a given 
value of the KGR. This can be indeed an advantage, also considering 
that the state-of-the-art technology exploiting PNR detectors cannot 
achieve the overall quantum efficiency of homodyne detectors.
In practice, homodyne setups may easily exhibit quantum efficiencies 
larger than $0.8$, whereas customary PNR detectors are about $0.5$ 
\cite{BinaOlivares} or less. Nevertheless, there exist photon-number-resolving
techniques based on transition-edge sensors which allows to obtain a much
higher efficiency, $\approx 0.9$ or higher \cite{TES1,TES2}.
\par
Further investigation is expected upon the design
of scheme for distilling a secret key after having run the protocol
and the investigation of performances for PNR detector based
protocols involving not just simple homodyne detection at the receiver,
but also heterodyne detection \cite{het1,het2,het3,optGauss,optGauss:2}.
In this view, our results pave the way for further developments in this 
promising field of quantum technology.

\section*{Acknowledgments}
We thank A.~Allevi and M.~Bondani for stimulating discussions,
A.~Leverrier for useful comments and M.~A.~C.~Rossi
for the assistance in the numerical calculations.
\vfill


\end{document}